\documentclass[aps,pra,preprint,showpacs,superscriptaddress,amsfonts,floatfix]{revtex4}
\usepackage{graphics}

\usepackage[dvips]{graphicx}
\usepackage{amsmath}

\newcommand{\Tr}{\mathop{\mathrm{Tr}}}

\begin{document}

%
%

\title{Exact spectral decomposition of a time-dependent one-particle reduced density matrix}

%
%
%
\author{I. Nagy}
\affiliation{Department of Theoretical Physics,
Institute of Physics, \\
Budapest University of Technology and Economics, \\ H-1521 Budapest, Hungary}
\affiliation{Donostia International Physics Center, P. Manuel de
Lardizabal 4, \\ E-20018 San Sebasti\'an, Spain}
\author{J. Pipek}
\affiliation{Department of Theoretical Physics,
Institute of Physics, \\
Budapest University of Technology and Economics, \\ H-1521 Budapest, Hungary}
\author{M. L. Glasser}
\affiliation{Department of Physics, Clarkson University, Potsdam,\\
New York 13699-5820, USA}
\affiliation{Donostia International Physics Center, P. Manuel de
Lardizabal 4, \\ E-20018 San Sebasti\'an, Spain}

\date{\today}
\begin{abstract}

We determine the exact time-dependent non-idempotemt one-particle reduced density matrix and its spectral decomposition for a harmonically confined two-particle correlated one-dimensional system when the interaction terms in the Schr\"odinger Hamiltonian are changed abruptly. Based on this matrix in coordinate space we derive
a precise condition for the equivalence of the purity and the overlap-square of the correlated and non-correlated
wave functions as the system evolves in time. This equivalence holds only if the interparticle interactions are affected, while the confinement terms are unaffected within the stability range of the system. 
Under this condition we also analyze
various time-dependent measures of entanglement and demonstrate that, depending on the magnitude of 
the changes made in the Schr\"odinger Hamiltonian,
periodic, logarithmically incresing or constant value behavior of the von Neumann entropy can occur.

\end{abstract}


\maketitle

\section{Motivation}

A quantum quench is an abrupt change in the state of a system due to changes in the potential energy terms 
of its Schr\"odinger Hamiltonian. For an intarcting system 
such a quantum quench can occur in the external confinement term, or in the interparticle terms. 
Both seem to be feasible in modern experiments \cite{Serwane11}
on trapped systems. Since, in general, the {\it interparticle} interaction is responsible
for dynamic correlation, beyond any statistics-mediated (exchange) correlation, 
its tuning is of special interest. 

Equally importantly, advances in optical trapping of cold atoms have allowed for an unprecedented 
manipulation over the size of these quantum systems such that the {\it number} of atoms 
beeing trapped can be precisely specified \cite{Wenz13}.
Clearly, exactly solvable entangled two-particle models could serve as benchmark systems
\cite {Brics13,Keller16}
to test the capability of approximate many-body methods in the time domain. 
Recently, motivated by the obvious theoretical interest on the fine details of informations, 
proposals have been made \cite {Demler12,Islam15} about measuring entanglement
in cold atom systems based on the interference between copies.

The present paper is organized as follows. The next, introductory Section is devoted to a summary of the
stationary model in order to provide a useful background. Section III contains the 
time-dependent extension. The particular case where the abrupt change is made
in the interparticle interaction is analyzed in details. Atomic units are used.

\section{Summary on the time-independent case}

Motivated by the really challenging interplay among Hamiltonian-based quantities and information-theory-based measures, we first summarize the 
stationary (i.e., time-independent) background to our present study. 
As our unquenched system, we take the interacting model system 
\cite{Heisenberg26,Moshinsky68} first introduced by Heisenberg as one of the really simplest ("Das denkbar einfachste Mehrk\"orperproblem") many-body models
\begin{equation}
\hat{H}(x_1,x_2)\, =\, -\, \frac{1}{2}\left(\frac{d^2}{dx_1^2}\, +
\frac{d^2}{dx_2^2}\right) +\frac{1}{2}\, \omega_0^2({x}_1^2+{x}_2^2)
-\frac{1}{2}\, \lambda\, \omega_0^2({x}_1-{x}_2)^2,
\end{equation}
where $\lambda$ measures the strength of the interparticle interaction energy in terms of $\omega_0^2$. 
The repulsive ($r$) and attractive ($a$) interactions
refer to $\lambda\equiv{\Lambda_r\in{[0,0.5]}}$ and $\lambda\equiv{\Lambda_a<0}$. The restricted range for $\Lambda_r$ 
will be clarified below. Physically, for $\Lambda_r>0.5$, both interacting particles 
cannot both remain in the confining external field.

Introducing standard normal coordinates $X_1\equiv{(x_1+x_2)/\sqrt{2}}$ and $X_2\equiv{(x_1-x_2)/\sqrt{2}}$,
one can easily rewrite the unperturbed Hamiltonian into the form 
\begin{equation}
\hat{H}(X_1,X_2)\, =\, -\, \frac{1}{2}\left(\frac{d^2}{dX_1^2}\, +
\frac{d^2}{dX_2^2}\right) +\frac{1}{2}\, \omega_1^2\, X_1^2 + \frac{1}{2}\, \omega_2^2\, X_2^2,
\end{equation}
where  $\omega_1\equiv{\omega_0}$ and $\omega_2\equiv{\omega_0\sqrt{1-2\lambda}}$ denote the 
frequencies of the independent normal modes. 
Based on Eq.(2), the normalized ground-state wave function $\Psi(X_1,X_2)$ is the product
\begin{equation}
\Psi(X_1,X_2)\, =\, \left(\frac{\omega_1}{\pi}\right)^{1/4}\exp\left[-\frac{1}{2}\omega_1\, X_{1}^2\right]\, 
\left(\frac{\omega_2}{\pi}\right)^{1/4}\exp\left[-\frac{1}{2}\omega_2\, X_{2}^2\right]. 
\end{equation}
We stress at this point that the price of this normal-mode transformation is that one loses
the intuitive physical picture
of real particles and, instead, operates with {\it effective} particles representing the transformed
coordinates. Since their frequencies are different one can not reinterpret them as quasiparticles,
at least within the famous Landau's picture.

Before turning to the reduced one-particle density matrix, a key quantity in our analysis, we introduce
here a more conventional measure
of interparticle correlation by calculating the overlap-{\it square} $O_1(\lambda)=|<\Psi(X_1,X_2,\lambda)|\Psi(X_1,X_2,\lambda=0)>|^2$.
The result becomes
\begin{equation}
O_1(\lambda)\, =\, \frac{2\sqrt{\omega_1\omega_2}}{\omega_1+\omega_2}.
\end{equation}
%
With the separated form for our model one obtains the same expression, 
$Q_1(\lambda)=Q_2(\lambda)$, from an other
overlap defined by interchanging (formally $x_2\Rightarrow{-x_2}$) the 
coordinates of {\it effective} particles in Eq.(3), i.e., from 
$O_2(\lambda):=|<\Psi(X_1,X_2,\lambda)|\Psi(X_2,X_1,\lambda)>|$.

By rewriting the wave function $\Psi(X_1,X_2)$ in terms of physical coordinates $x_1$ and $x_2$, we
determine the one-matrix $\Gamma_1(x_1,x_2)$ via the following nonlinear mapping
\begin{equation}
\Gamma_1(x_1,x_2)\, =\, \int_{-\infty}^{\infty}\, dx_3\,
\Psi^{*}(x_1,x_3)\, \Psi(x_2,x_3). 
\end{equation}
From this, we arrive at a very informative, Jastrow-like \cite{Ochi17}, representation
\begin{equation}
\Gamma_1(x_1,x_2)=\phi_s(x_1)\, \phi_s(x_2) 
\times{e^{-D[(x_1-x_2)/\sqrt{2}]^2}}
\end{equation}
where, with $\omega_s\equiv{2\omega_1\omega_2/(\omega_1+\omega_2})$, we introduced the following abbreviations
\begin{equation}
\phi_s(x)=\left[\frac{\omega_s}{\pi}\right]^{1/4}\,
e^{-\frac{1}{2}\omega_s\, x^2}     
\end{equation}
\begin{equation}
D\, =\, \frac{1}{4}\, \frac{(\omega_1-\omega_2)^2}
{\omega_1+\omega_2}\, \geq{0}. 
\end{equation}

The diagonal of $\Gamma_1(x_1,x_2)$ gives the one-particle 
probability density, $n(x)=\Gamma(x,x)$, of unit-norm, which is the basic quantity
of the mean-field Density Functional Theory (DFT). However, if we neglect the role of the relative coordinate
by taking $D=0$ in Eq.(6), we get an idempotent one-matrix \cite{Dreizler90}. 
In DFT a density-optimal auxiliary orbital, $\phi_s(x)$, is occupied with unit probability. 
The double Fourier transform \cite{Dreizler90} of the one-dimensional one-matrix in Eq.(6) 
gives the momentum-dependent one-matrix. After integrations we get
\begin{equation}
\Gamma_1(k_1,k_2)=\frac{1}{\sqrt{\pi(\omega_s+2D)}}\, e^{-\frac{1}{2}(k_1^2+k_2^2)\frac{\omega_s+D}
{\omega_s(\omega_s+2D)}}\, e^{+\frac{Dk_1k_2}{\omega_s(\omega_s+2D)}}. \nonumber
\end{equation}
The diagonal component of $\Gamma_1(k,k)$ is the normalized one-body momentum distribution function 
and its application gives the {\it exact} kinetic energy of the system:
$<K>=(1/4)(\omega_s+2D)=(1/4)(\omega_1+\omega_2)$. Clearly, by using the virial theorem for 
bounded systems with harmonic interactions, one gets the exact ground-state
energy as well. Since one has access to $\Gamma_1(k,k)$ 
by Compton scattering, it is an observable encoding information on entanglement.

Next we calculate the purity \cite{Pipek09} defined by
\begin{equation}
\Pi(\lambda)\, :=\, \Tr[\Gamma^2_1]\, = \, \int\, \Gamma_1^2(x,x)\, dx,
\end{equation}
where, the operator-square $\Gamma_1^2(x_1,x_2)$ is written
\begin{equation}
\Gamma_1^2(x_1,x_2)\, =\, \int\, \Gamma_1(x_1,x_3)\Gamma_1(x_3,x_2)\, dx_3.
\end{equation}
The direct calculation of the purity, based on Eqs.(9-10) with Eq.(6), results in
\begin{equation}
\Pi(\lambda)\, =\, \frac{1}{(1+2D/\omega_s)^{1/2}}\, \equiv{\frac{2\sqrt{\omega_1\omega_2}}{\omega_1+\omega_2}}\,
=\, \frac{\omega_s}{\sqrt{\omega_1\omega_2}}\, =\, O_1(\lambda).
\end{equation}
This is a remarkable equality. It says that two measures of probabilistic nature are equal
in our entangled model system. If one of them, 
say the overlap, could be accessible experimentally \cite{Demler12,Islam15}, 
we could characterize the other one as well.

The purity rests on an operator-{\it square}, and is calculable directly from knowledge of the coordinate-representation of the one-particle reduced density matrix. However, for a compact spectral analysis \cite{Calabrese15} of correlation one needs $\Tr[\Gamma_1^q]$ for noninteger $q$ values as well. 
Based on them one can calculate R\'enyi's and von Neumann's entropies \cite{Renyi70}.
The possibility of point-wise \cite{Riesz55} direct decomposition of a two-variable function rests on the mathematical observation that 
Mehler's formula \cite{Erdelyi53,Robinson77,Glasser13} gives
\begin{equation}
(\bar{\omega}/\pi)^{1/2}\,  e^{- \frac{\bar{\omega}}{2}  \left(\frac{1+Z^2}{1-Z^2}\right) (x_1^2+x_2^2)}\,  e^{\bar{\omega} \frac{2\, Z}{1-Z^2}\, x_1 x_2}=
\sum_{k=0}^{\infty} (1-Z^2)^{1/2}\, Z^k\, \phi_k(\bar{\omega},x_1)\, \phi_k(\bar{\omega},x_2),
\end{equation}
where the parameter $Z\in{[0,1]}$, and $x_i\in{(-\infty,\infty)}$. The $\phi_k(\bar{\omega},x)$ decomposition-functions form a 
complete set of orthonormal eigenfunctions of a one-dimensional harmonic oscillator with potential energy 
$\bar{\omega}^2(x^2/2)$ in the Schr\"odinger wave equation and are given by
\begin{equation}
\phi_k(\bar{\omega},x)\, = \,
\left(\frac{\bar{\omega}}{\pi}\right)^{1/4}\frac{1}{\sqrt{2^k\,
k!}}\, e^{-\frac{1}{2}\bar{\omega}\, x^2}\,
H_k(\sqrt{\bar{\omega}}x).
\end{equation}
Comparison of exponentials in Eq.(6) and Eq.(12) results in the two constraints 
\begin{equation}
(\omega_s+D)\, =\, \bar{\omega}\, \frac{1+Z^2}{1-Z^2}
\end{equation}
\begin{equation}
D\, =\, \bar{\omega}\, \frac{2Z}{1-Z^2}.
\end{equation}
One can solve the algebraic equations easily for $Z$ and $\bar{\omega}$ in terms of $D$ and $\omega_s$. We get
\begin{equation}
Z(\lambda)\, =\, \frac{\sqrt{1+2D/\omega_s}-1}{\sqrt{1+2D/\omega_s}+1}\, =\, 
\left(\frac{\sqrt{\omega_1}-\sqrt{\omega_2}}{\sqrt{\omega_1}+\sqrt{\omega_2}}\right)^2 
\end{equation}
\begin{equation}
\bar{\omega}\, =\, \omega_s\sqrt{1+2D/\omega_s}\, \equiv{\sqrt{\omega_1\omega_2}}.
\end{equation}
It follows from Eq. (16) that $Z(\Lambda_a)=Z(\Lambda_r)$, i.e., there is a duality
\cite{Pipek09,Rolf14}
under the constraint of $\Lambda_a=-\Lambda_r/(1-2\Lambda_r)$. Thus, under this constraint
at stationary condition, probabilistic measures alone can not reproduce 
the {\it sign} of the interparticle interaction.

Since $\omega_s=\bar{\omega}(1-Z)/(1+Z)$ from Eqs.(14-15), 
we obtain a point-wise, closed-shell-like expansion for the one-particle reduced density matrix
\begin{equation}
\Gamma_1(x_1,x_2,\bar{\omega})\, =\, \sum_{k=0}^{\infty}\, P_k(Z)\,
\phi_k(\bar{\omega},x_1)\, \phi_k(\bar{\omega},x_2),
\end{equation}
where the occupation numbers of the so-called natural orbitals, $\phi_k(\bar{\omega},x)$, are 
\begin{equation}
P_k\, =\, (\omega_s/\bar{\omega})^{1/2}\, (1-Z^2)^{1/2} Z^k\, =\, (1-Z) Z^k
\end{equation}
and we have $\sum_{k=0}^{\infty}P_k=1$. 
In the knowledge of occupation numbers, i.e., the eigenvalues of the one-matrix \cite{Srednicki93},
one can calculate R\'enyi's entropies \cite{Renyi70} for $0<q<\infty$, 
since in our case $(P_k)^q=(1-Z)^q(Z^q)^k$.
Thus, a desirable \cite{Calabrese15} spectral analysis of information-theoretic measures 
becomes feasible. For instance, as a useful check, we get
\begin{equation}
\Pi(\lambda)\, =\, O_1(\lambda)\, =\, \frac{1}{\sqrt{1+ 2D/\omega_s}}\equiv{\sum_{k=0}^{\infty} (P_k)^2} = \frac{1-Z(\lambda)}{1+Z(\lambda)} = 
\frac{\omega_s}{\bar{\omega}} = \frac{2(1-2\lambda)^{1/4}}{1+\sqrt{1-2\lambda}}\leq{1} \nonumber
\end{equation}
This series exhibits useful connections between physical and auxiliary variables.

\newpage

\section{Results for the time-dependent case}

Motivated by the remarkable experimental possibilities, outlined in the first Section 
on trapped systems with controllable numbers of constituents, we suppose abrupt changes 
are made in Heisenberg's Hamiltonian at $t=0$ and have for $t>0$ for the new ($n$) Hamiltonian
\begin{equation}
\hat{H}_n(x_1,x_2)\, =\, -\, \frac{1}{2}\left(\frac{d^2}{dx_1^2}\, +
\frac{d^2}{dx_2^2}\right) +\frac{1}{2}\, \Omega_0^2({x}_1^2+{x}_2^2)
-\frac{1}{2}\, \lambda'\, \Omega_0^2({x}_1-{x}_2)^2.
\end{equation}
$\Omega_0=0$, is considered a complete quench. This
results in a {\it free} propagation \cite{Kagan96,Rubio12} of the initially entangled system.
As above, we use this Hamiltonian in its separated form
\begin{equation}
\hat{H}_n(X_1,X_2)\, =\, -\, \frac{1}{2}\left(\frac{d^2}{dX_1^2}\, +
\frac{d^2}{dX_2^2}\right) +\frac{1}{2}\, \Omega_0^2\, X_1^2 + \frac{1}{2}\, \Omega_2^2\, X_2^2,
\end{equation}
where $\Omega_2=\Omega_0\sqrt{1-2\lambda'}$. The particular choice of $\Omega_0=\omega_0$ 
and $\lambda'\neq{0}$ in $\hat{H}_n(x_1,x_2)$, 
but taking $\lambda=0$ in $\hat{H}(x_1,x_2)$, refers to entanglement 
{\it production} addressed recently \cite{Yukalov15}.

As we have a separated form for $\hat{H}_n(X_1,X_2)$ in the time-dependent Schr\"odinger equation for $t>0$, 
we proceed by propagating \cite{Haar75}
independently both normalized stationary normal ($j=1,2$) modes,
$\psi_1(X_1,\omega_1)$ and $\psi_2(X_2,\omega_2)$ from $\Psi(X_1,X_2)=\psi_1(X_1)\psi_2(X_2)$ of Eq.(3), 
by using the corresponding ($\Omega_j^2\geq{0}$, i.e., the $\lambda'>0.5$ explosion-case is excluded) propagators
\begin{equation}
G_j(X_j,X_j',\Omega_j,t)\, =\, \left(\frac{\Omega_j}{2{\pi}\, i \sin{\Omega_j t}}\right)^{1/2}
\exp\left\{\frac{i\Omega_j}{2\sin{\Omega_j t}}\, [(X_j^2 + {X_j'}^2)\cos{\Omega_j t} - 2X_j {X_j}']\right\} \nonumber
\end{equation}
which reproduces the free-propagation case when one first takes 
$\Omega_j\rightarrow{0}$ at fixed $t$. 
With this textbook $G_j(t)$, we have to perform the convolution 
\begin{equation}
\psi_j(X_j,t)\, =\, \int_{-\infty}^{\infty}\, d{X_j}'\, G_j(X_j,{X_j}',\Omega_j,t)\,
\psi_j({X_j}'),
\end{equation}
to derive $\Psi_n(X_1,X_2,t)=\psi_1(X_1,t)\psi_2(X_2,t)$. We arrive at the normalized solution
\begin{equation}
\psi_j(X_j,t)\, =\, \left(\frac{B_j}{\pi}\right)^{1/4}\,  
\exp\left[-\frac{1}{2}{X_j}^2 B_j(1-i\, C_j)\right]
\end{equation}
where the coefficients, $B_j(\omega_j,\Omega_j,t)$ and $C_j(\omega_j,\Omega_j,t)$, are given by
\begin{equation}
B_j(t)\, =\, 
\omega_j\, \frac{{\Omega_j}^2}{{\Omega_j}^2\cos^2{\Omega_j t} + {\omega_j}^2\sin^2{\Omega_j t}} \nonumber 
\end{equation} 
\begin{equation}
C_j(t)\, =\, \frac{1}{2}\, \frac{{\omega_j}^2-{\Omega_j}^2}{\omega_j\Omega_j}\, \sin(2\Omega_j t) \nonumber
\end{equation} 

In the complete-quench case we get $C_j(t)=\omega_j\, t$ and $B_j(t)=\omega_j/(1+\omega_j^2\, t^2)$, and thus
$B_j(t)[1+C_j^2(t)]=\omega_j$. Here one has a simple free expansion without oscillation. 
In other quenches, we expect characteristicly oscillating wave functions and 
corresponding probability measures. 
However, and this is very important from physical point of view,
the period of these oscillations will now depend on the sign of the interparticle interactions.
For the overlap-square defined by $O_1(t)=|<\Psi_n(X_1,X_2,\lambda',t)|\Psi(X_1,X_2,\lambda=0)>|^2$, one gets
\begin{equation}
O_1(t)\, =\, \frac{2\sqrt{\omega_0\, B_1}}{[(\omega_0 + B_1)^2+(B_1\, C_1)^2]^{1/2}}\,
\frac{2\sqrt{\omega_0\, B_2}}{[(\omega_0 + B_2)^2+(B_2\, C_2)^2]^{1/2}},
\end{equation}
which reproduces the stationary result in Eq.(4) at $t=0$, $\Omega_0=\omega_0$ and $\lambda'=\lambda$. We stress that we evaluate this overlap with the stationary {\it noninteracting} state, as at Eq.(4).

By rewriting the evolving wave function $\Psi_n(X_1,X_2,t)$ in terms of original coordinates $x_1$ and $x_2$, we
determine the one-matrix $\Gamma_1(x_1,x_2,t)$ from the following nonlinear mapping
\begin{equation}
\Gamma_1(x_1,x_2,t)\, =\, \int_{-\infty}^{\infty}\, dx_3\,
\Psi_n^{*}(x_1,x_3,t)\, \Psi_n(x_2,x_3,t), 
\end{equation}
After a long, but quite straightforward, calculation we obtain
\begin{equation}
\Gamma_1(x_1,x_2,t)=\phi_s(x_1,t)\, \phi_s^{*}(x_2,t)\,
e^{-\frac{1}{2}D(t)(x_1-x_2)^2},
\end{equation}
in which we introduced a time-dependent [c.f., Eq.(7)] auxiliary function
\begin{equation}
\phi_s(x,t)=\left[\frac{\omega_s(t)}{\pi}\right]^{1/4}\,
e^{-\frac{1}{2}\omega_s(t)\, x^2}\, e^{i\frac{1}{2}\omega_s(t)\, C_s(t)\, x^2}      
\end{equation}
where $\omega_s(t)=2B_1(t)B_2(t)/[B_1(t)+B_2(t)]$ and $C_s(t)=[C_1(t)+C_2(t)]/2$. Clearly, 
the individual normal-mode characters ($B_j$ and $C_j$) influence the auxiliary $\phi_s(x,t)$
state via their differently {\it weighted} forms. The differences of characters will appear in
\begin{equation}
D(t)\, =\, \frac{1}{4}\, \frac{[B_1(t)-B_2(t)]^2
+[B_1(t)C_1(t) - B_2(t)C_2(t)]^2}{B_1(t)+B_2(t)}\, \geq{0}. 
\end{equation}

By taking $D(t)\equiv{0}$, i.e., neglecting the role of the relative coordinate, one arrives at the 
idempotent one-matrix of Time-Dependent Density-Functional Theory \cite{Ullrich12}). 
The diagonal ($x_1=x_2=x$) part of 
$\Gamma_1(x_1,x_2,t)$ gives the exact probability density $n(x,t)=[\phi_s(x,t)]^2$, the basic variable
of TDDFT, a mean-field theory. The other variable in a physically consistent approximation in TDDFT
is the probabilty current $j(x,t)$, needed in the fundamental continuity equation $\partial_t n(x,t)+\partial_x j(x,t)=0$. The current is defined, and is given by
\begin{equation}
j(x,t)\, :=\, Re\, \int_{-\infty}^{\infty} \Psi^{*}_n(x,x',t)\, \frac{\partial}{i\, \partial x}\, 
\Psi_n(x,x',t)\, dx'\, = n(x,t)[x\, \omega_s(t)]\, C_s(t). \nonumber
\end{equation}
It is easy to show, after substitution, that in the complete-quench case we get 
$2D(t)/\omega_s(t)=2D/\omega_s$ for $t>0$, heralding 
rigidity in the entropic-measures for free-propagation. Such a memory is expected on physical grounds.
Furthermore, as we will discuss for the most important case of abrupt
changes in the {\it interparticle} interaction,
there could be finite $t$ values at which we get $[2D(t_1)/\omega_s(t_1)]=2D/\omega_s$ or 
$[2D(t_2)/\omega_s(t_2)]=0$, both periodically.

The one-matrix in Eq.(26) is  self-adjoint $\Gamma_1(x_1,x_2,t)=\Gamma_1^{*}(x_2,x_1,t)$ and positive.
By its application in the defining Eqs.(9-11) we get directly
\begin{equation}
\Pi(t)\, =\, \Tr\, [\Gamma^2_1(t)]\, =\, \frac{1}{[1+2D(t)/\omega_s(t)]^{1/2}}.
\end{equation}
Using this at $t=0$, where $B_j(t=0)=\omega_j$ and $C_j(t=0)=0$, 
the stationary result of the previous Section is recovered, since $\sqrt{1+2D/\omega_s}=[1+Z(\lambda)]/[1-Z(\lambda)]$. 
Thus, in the complete quench case the purity remains constant. However, the overlap $O_1(t)$ tends to zero, as expected, and for free propagation $O_1(t\rightarrow{\infty})\simeq{\sqrt{\omega_0/\omega_2}\, [2/(t\omega_0)]^2}=4/(\omega_0\bar{\omega}t^2)$. The overlap $O_2(t)$,
defined at Eq.(4) for the stationary case, behaves with respect to time as
\begin{equation}
O_2(t):=|<\Psi^{*}(X_1,X_2,\lambda,t)|\Psi(X_2,X_1,\lambda,t)>|\equiv{\frac{1}{[1+2D(t)/\omega_s(t)]^{1/2}}}=\Pi(t). \nonumber
\end{equation}
Thus, in the two-body model investigated, the interchange of coordinates of {\it effective} particles 
can result in $O_1(t)\neq{O_2(t)}$ for $t\neq{0}$. A related analysis will be given at Eq.(34), below.

Following closely our arguments behind a point-wise decomposition of the stationary 
one-matrix in the previous Section, 
we turn now to the time-dependent case.
Thus, we introduce time-dependent $\bar{\omega}(t)$ and $Z(t)$, via the two constraints 
\begin{equation}
[\omega_s(t)+D(t)]\, =\, \bar{\omega}(t)\frac{1+Z^2(t)}{1-Z^2(t)} \nonumber
\end{equation}
\begin{equation}
D(t)\, =\, \bar{\omega}(t)\frac{2Z(t)}{1-Z^2(t)} \nonumber
\end{equation}
One can solve these equations for $\bar{\omega}(t)$ and $Z(t)$ in terms of $D(t)$ and $\omega_s(t)$
to obtain
\begin{equation}
Z(t)\, =\, \frac{\sqrt{1+2D(t)/\omega_s(t)}-1}{\sqrt{1+2D(t)/\omega_s(t)}+1}
\end{equation}
\begin{equation}
\bar{\omega}(t)\, =\, \omega_s(t)\sqrt{1+2D(t)/\omega_s(t)}.
\end{equation}
In terms of these solutions we get the point-wise spectral decomposition 
\begin{equation}
\Gamma_1(x_1,x_2,t)\, =\, \sum_{k=0}^{\infty}\, P_k(t)
\phi_k(x_1,t)\, \phi_k^{*}(x_2,t),
\end{equation}
where $P_k(t)=([1-Z(t)]\, [Z(t)]^k$, and $\phi_k(x,t)$ are the so-called natural orbitals
\begin{equation}
\phi_k(x,t)\, =\,
\left[\frac{\bar{\omega}(t)}{\pi}\right]^{1/4}
\left[\frac{1}{\sqrt{2^k\, k!}}\,
e^{-\frac{1}{2}\bar{\omega}(t)x^2}\,
H_k(\sqrt{\bar{\omega}(t) x})\right]\, e^{i\, \frac{1}{2}\omega_s(t)\, C_s(t)\, x^2}.
\end{equation}

Knowing $P_k(t)$, one can easily calculate different time-dependent  
measures for entanglement such as $\Pi(t)$, R\'enyi's [$S_R(q,t)$] 
and von Neumann [$S_N(t)=S_R(q\rightarrow{1},t)$] entropies 
\begin{equation}
\Pi(t)\, =\, \frac{1-Z(t)}{1+Z(t)} \nonumber
\end{equation}
\begin{equation}
S_R(q,t)\, =\, \frac{1}{1-q}\,
\ln\frac{[1-Z(t)]^q}{1-[Z(t)]^q} \nonumber
\end{equation}
\begin{equation}
S_N(t)=-\left[q^2 \frac{d}{dq} \left(\frac{1-q}{q} S_R(q,t)\right)\right]_{q=1}\,
=\, -\ln[1-Z(t)]\, -\frac{Z(t)}{1-Z(t)} \ln[Z(t)]. \nonumber
\end{equation}

Remember, that we obtained for the stationary case the remarkable equality in Eq.(11), 
$\Pi(\lambda)=O_1(\lambda)$. Is there a similar, possibly very useful, equality in the time-dependent case? 
The answer is yes, {\it if and only if} we make abrupt changes in the interparticle
interaction only. For such changes, where $\Omega_0=\omega_0$ but $\lambda'\neq{\lambda}$, instead of Eq.(28) we have
\begin{equation}
D(t)\, =\, \frac{1}{4}\, \frac{[\omega_0-B_2(t)]^2
+[B_2(t)C_2(t)]^2}{\omega_0+B_2(t)}\, \geq{0}. \nonumber
\end{equation}
since $B_1(t)=\omega_0$ and $C_1(t)=0$. From Eq.(24) [$\omega_s(t)=2\omega_0B_2(t)/(\omega_0+B_2(t)$] we obtain
\begin{equation}
\Pi(t)=O_2(t)={\frac{1}{[1+2D(t)/\omega_s(t)]^{1/2}}}
\equiv{\frac{2\sqrt{\omega_0\, B_2}}{[(\omega_0 + B_2)^2+(B_2\, C_2)^2]^{1/2}}}= O_1(t) .
\end{equation}
We stress, that this is true when we manipulate, {\it solely}, the interparticle interaction. 
Remarkably, such a (gedanken) case was addressed first by Einstein, Podolsky, and Rosen (EPR) 
in \cite{Einstein35}. Now, this seems to be feasible in modern experiments on trapped systems. 

For our EPR-case, we derive expressions exhibiting the role of $\lambda\rightarrow{\lambda'}$ changes.
%
\begin{equation}
1+\frac{2D(t)}{\omega_s(t)} = \left(1+\frac{2D}{\omega_s}\right)+\frac{1}{\Omega_2}\left(\frac{\Omega_2^2 -\omega_1^2}{\omega_1\, \Omega_2}\right)\, 
\frac{\Omega_2^2-\omega_2^2}{4\, \omega_2}\,
\sin^2(\Omega_2 t). 
\end{equation}
Based on this equation we arrive, after substitutions, at the following
\begin{equation}
\left[\frac{1+Z(\lambda,t)}{1-Z(\lambda,t)}\right]^2 = \left[\frac{1+Z(\lambda)}{1-Z(\lambda)}\right]^2 + 
\frac{\lambda'(\lambda'-\lambda)}{\sqrt{1-2\lambda}(1-2\lambda')}
\sin^2(\omega_0 t \sqrt{1-2\lambda'}).
\end{equation}
which shows that at $t\rightarrow{0}$ the stationary result will vary quadratically in time.
In that limit, the associated changes in entropic measures will exhibit a similar, $\propto{t^2}$, scaling. 

Notice that the long-time behavior in the repulsive case needs some care. When we tune, from below,
the repulsive $\lambda<0.5$ coupling to its critical limit $\lambda'\rightarrow{0.5}$, we arrive at $\Omega_2\rightarrow{0^{+}}$ in
the parameters given at Eq.(23), i.e., we have $B_2(t)=\omega_2/[1+C_2^2(t)]$ and $C_2(t)=\omega_2\, t$. 
The corresponding overlap $O_1(t)$, and thus $\Pi(t)$, tend to zero at $t\rightarrow{\infty}$. In that asymptotic 
limit for critical coupling ($\lambda'\rightarrow{0.5}$) our $Z(t)$ tends to unity as 
\begin{equation}
Z(t\rightarrow{\infty})\, \simeq{\, 1 - \frac{2(1-2\lambda)^{1/4}}{\sqrt{0.5(0.5-\lambda)}}\, \frac{1}{t\, \omega_0}}\,
=\, 1 - \frac{4}{t\, \omega_0(1-2\lambda)^{1/4}}.  \nonumber
\end{equation}
This gives for the sign-dependent, limiting behavior in the von Neumann entropy
\begin{equation}
S_N(t\,\bar{\omega}\gg{1})\, \simeq{\ln(t\, \bar{\omega})}. \nonumber
\end{equation}
Slow, i.e., logarithmic, growth of entanglement is a hot topic in 
nonequilibrium many-body systems \cite{Moore12,Serbyn13} as well.
Because of such similarity, we are tempted to think in terms of a kind of universality. 
If this turns out to be true rigorously, then critical transitions in trapped and solid-state
disordered systems could get a common probabilistic characterization.

Based on Eq.(36), we turn to {\it illustrative} cases which are exhibited in Figures 1-2. 
We stress, that only the EPR-like situation, i.e., a $\lambda\rightarrow{\lambda'}$ change,
is considered.
When $\lambda'=0$, i.e., when we turn off abruptly the interparticle interactions, 
there are no changes in entropies and purity since $Z(\lambda,t)=Z(\lambda)$.
Similar, memory-like character in entanglement was found recently in \cite{Ikeda15}
on pre-thermalization in a one-dimensional Bose system.
There, further arguments are given about a possible generic phenomenon of keeping entanglement.

In Figuere 1 showing $\Pi(t)$, we take $\Lambda_r=0.49$ as before-quench value for repulsive coupling ($\lambda$).
The dotted curve refers to a change $\lambda\rightarrow{\lambda'}$ where, after quench, 
one has $\Lambda_r'=(1-\sqrt{1-2\Lambda_r})/2$. Thus we tune the interaction in order to 
get $\Omega_2\equiv{\bar{\omega}}$, i.e., the frequency in stationary natural orbitals. 
As expected for the physical consequence of minimization ($m$) on the rhs of Eq.(36), 
we get $Z(t_m)=0$ and thus exactly zero entropies, at times $t$ where $t_m\Omega_2=t_m\bar{\omega}=(2m+1)\pi/2$.
The solid curve refers to a change prescribed by $\lambda'=1.01\lambda$, i.e., to a $1\%$ enhancement in coupling.

\begin{figure}
\scalebox{0.5}[0.5] {\includegraphics{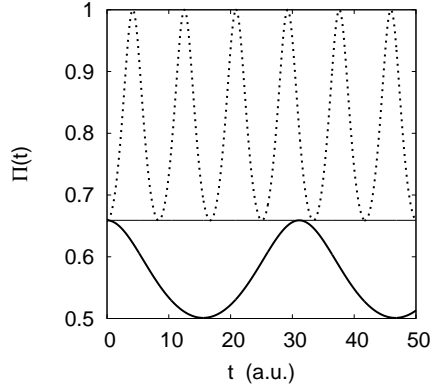}} \caption
{Time-dependent purity $\Pi(t)$ in the repulsive ($r$) case as a function of the time $t$, which is measured in atomic units $(a.u.)$. 
Eq. (36) with fixed $\omega_0=1$ is used. The stationary system is characterized by $\Lambda_r=0.49$. 
The dotted and solid curves refer to $\Lambda_r'=[(1-\sqrt{1-2\Lambda_r})/2]<\Lambda_r$, 
and a small enhancement $\Lambda_r'=1.01\Lambda_r$, respectively. 
See the text for further details.
\label{figure1}}
\end{figure}

\begin{figure}
\scalebox{0.5}[0.5] {\includegraphics{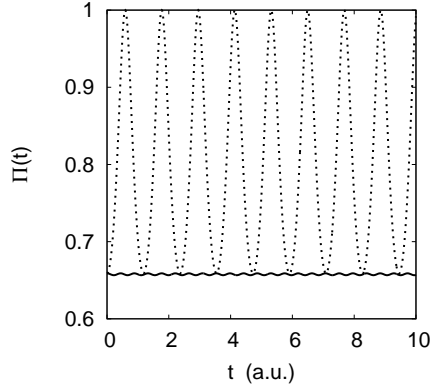}} \caption
{Time-dependent purity $\Pi(t)$ in the attractive ($a$) case as a function of the time $t$, which is measured in atomic units $(a.u.)$. 
Eq. (36) with fixed $\omega_0=1$ is used. The stationary system is characterized by $\Lambda_a=-24.5$. 
The dotted and solid curves refer to $\Lambda_a'=[(1-\sqrt{1-2\Lambda_r})/2]$, 
and a small increase $\Lambda_a'=1.01\Lambda_a$, respectively. 
See the text for further details.
\label{figure2}}
\end{figure}

Figure 2 on $\Pi(t)$ is devoted to the attractive case. We use $\Lambda_a=-24.5$, at which $Z(\Lambda_r)=Z(\Lambda_a)$, i.e.,
we have, before changes, a duality in entropic measures. Similarly to Figure 1, dotted and solid curves refer to $\Lambda_a'=(1-\sqrt{1-2\Lambda_a})/2$
and $\lambda'=1.01\lambda$, respectively. Comparison of the illustrative Figures shows that, 
in the evolving cases, the oscillation-periods reflect the {\it sign} of interparticle interaction.
Oscillations in a time-dependent measure were found in \cite{Yukalov15}, for
an entanglement production where $\lambda=0$ initially, and $\lambda'\neq{0}$ for $t>0$. 
Oscillations in time appear in the Loschmidt echo in a numerically exacly treated, 
one-dimensional two-body harmonically confined system driven by adding an external localized impurity \cite{Keller16}.
Based on Eq.(36), we could easily characterize the case of $(\lambda=\Lambda_r)\rightarrow{(\lambda'=\Lambda_a)}$, i.e., 
the repulsive-to-attractive quench, addressed recently in \cite{Haque15}.

\section{Conclusions and comment}

Based on Heisenberg's two-particle model Hamiltonian which constitutes a cornerstone in such areas of physics, as the fundamental fields of correlated atoms and confined quantum matter, a detailed analysis of entanglement measures is performed using the exact decomposition of a time-dependent one-matrix obtained by abrupt changes in the coupling of the interparticle interaction energy. 
In the exact spectral representation of the one-matrix the natural orbitals and their 
occupation numbers depend on time, as expected. Using such representation a precise condition for 
the equivalence of purity and an overlap-{\it square} is derived for the interacting 
two-particle model in the time-dependent case. 

For couplings in the stability range our information-theoretic measures based on the occupation numbers oscillate in time.
Remarkably, recent work \cite{Dast16} found oscillations in purity of {\it two-mode} Bose-Einstein many-body system.
It was argued that such oscillation is measurable via the average contrast in interference experiments.
Comparison of attractive and repulsive cases shows that the stationary dual character of entropies may change.
In the time-evolving situation the originally equal entropies can be restored only at different times, periodically.
In the complete quench case the overlap-square tends to zero as the inverse second-power of time. 
At the stability limit for repulsive coupling we get, at $t\rightarrow{\infty}$, a logarithmic increase 
in the von Neumann entropy. This behavior resembles to the one found at transitions in disordered 
many-body systems \cite{Moore12,Serbyn13}. 

Exactly solvable two-particle models of controllably trapped systems can serve as
benchmarks in the challenging field dealing with encoded informations in correlated systems. Our 
analytic results on the time domain could, therefore, contribute to a general understanding
in that hot topic field. Further research, based on other two-body model systems with harmonic
confinement but with different interparticle intaraction, is desirable for a detailed
comparison and, as a main goal of such research, to provide a transferable knowledge.

Finally, based on our exact result on an entangled two-body model, we comment on the effective
potential, $V_s(x,t)$, in practical TDDFT. That mean-field method of electronic-structure calculations 
rests on auxiliary orbitals, $\phi_s(x,t)$, with integer occupation. 
Using the exact probabilistic inputs given at Eq.(28) 
for the density and current, and following the prescription detailed on page 105 of \cite{Ullrich12}
on a sophisticated {\it inversion}, the effective potential to be used in the associated 
Schr\"odinger-like equation for the auxiliary orbital becomes
\begin{equation}
V_s(x,t)=\frac{1}{2}\omega_s^2(t)x^2 - \frac{1}{2}\left[\sqrt{\omega_s(t)}
\frac{d^2}{dt^2}\left(\frac{1}{\sqrt{\omega_s(t)}}\right)\right]x^2, \nonumber
\end{equation}
in our case, upto a time-dependent constant. For the complete quench condition, 
one still would \cite{Schirmer07,Maitra15} get via such inversion
$V_s(x,t>0)\propto{x^2(\omega_s^2-\bar{\omega}^2)
/[1+(\bar{\omega}t)^2]^2}\neq{0}$, if $\lambda\neq{0}$.

However, by {\it starting} with results $\phi_s(x)$ and thus precise $n(x)$ of stationary DFT,
the initial ($t<0$) effective potential $V'_s(x)=(1/2)\omega_s^2\, x^2$ will change, after a complete quench,
to the expected $V'_s(x,t>0)=0$ form since in the above equation one has to use the corresponding
$\omega'_s(t)=\omega_s/[1+(\omega_s t)^2]$ instead of $\omega_s(t)$ based on weighting two different modes.
Thus, although $n'(x,t)\neq{n(x,t)}$ and $j'(x,t)\neq{j(x,t)}$ 
[but $\partial_t n'(x,t)+\partial_x j'(x,t)=0$] at such practical propagation from a pre-optimized state,
the physically desirable (at the complete quench made) character of effective interaction for $t>0$ is restored.

\begin{acknowledgments}
Two of us (I.N., M.L.G.) thank Professor  P. M. Echenique for the very warm hospitality 
at the DIPC. Useful discussions with Professors  B. D\'ora, Ch. Schilling, and G. Tak\'acs
on several aspects of the problem investigated in this work are acknowledged.
\end{acknowledgments}

\end{document}